\documentclass[preprint,12pt]{elsarticle}

\usepackage{graphicx}
\usepackage{amssymb}
\usepackage{arydshln}
\usepackage{wrapfig}
\usepackage{siunitx}
\usepackage{mathtools}
\usepackage{multirow}
\usepackage[table]{xcolor}
\usepackage{array}
\usepackage{appendix}
\usepackage{float}

\usepackage[
  colorlinks=true,
  linkcolor = blue,
  urlcolor  = blue,
 citecolor = blue,
  breaklinks,
  unicode
]{hyperref}

\usepackage{lineno}
\usepackage{rotating}
\usepackage{supertabular,booktabs}
\usepackage{longtable}
\usepackage{lscape}
\usepackage{geometry}
\usepackage{adjustbox}
\usepackage{natbib}
\usepackage[version=4]{mhchem}

\usepackage[english]{babel}
\usepackage{blindtext}
\newcolumntype{P}[1]{>{\centering\arraybackslash}p{#1}}

\makeatletter
\def\ps@pprintTitle{%
   \let\@oddhead\@empty
   \let\@evenhead\@empty
   \let\@oddfoot\@empty
   \let\@evenfoot\@oddfoot
}
\makeatother

\begin{document}


\begin{frontmatter}

\title{A backing detector for order-keV neutrons}

\author[1]{A. Biekert} 
 \author[4]{L.~Chaplinsky}
 \author[1]{C.W.~Fink} 
 \author[2]{M.~Garcia-Sciveres}
 \author[4]{W. C. Gillis}
\author[7,8]{W.\,Guo} 
\author[4]{S.A.~Hertel} 
 \author[6]{G.~Heuermann}
\author[2]{X. Li}
\author[1]{J.~Lin} 
\author[9]{R.~Mahapatra} 
\author[1,2]{D.N.~McKinsey} 
\author[4]{P.K.~Patel\corref{cor1}}
\author[6]{B.~Penning} 
\author[4]{H.D.~Pinckney} 
\author[9]{M.~Platt}
\author[1]{M.~Pyle} 
\author[1]{R.K.~Romani} 
\author[4]{A.~Serafin}
\author[1]{R.J.~Smith} 
\author[1]{B.~Suerfu} 
\author[1]{V.~Velan}
\author[3]{G.~Wang} 
\author[1]{Y.~Wang}
 \author[1]{S.L.~Watkins} 
 \author[6]{M.R.~Williams}

\address[1]{University of California Berkeley, Department of Physics, Berkeley, CA 94720, USA}
\address[2]{Lawrence Berkeley National Laboratory, 1 Cyclotron Rd., Berkeley, CA 94720, USA} 
\address[3]{Argonne National laboratory, 9700 S Cass Ave, Lemont, IL 60439, USA}
\address[4]{University of Massachusetts, Amherst Center for Fundamental Interactions and Department of Physics, Amherst, MA 01003-9337 USA} 
\address[5]{Walter  Burke  Institute  for  Theoretical  Physics, California  Institute  of  Technology,  Pasadena,  CA  91125,  USA}
\address[6]{University of Michigan, Randall Laboratory of Physics, Ann Arbor, MI 48109-1040, USA} 
\address[7]{Mechanical Engineering Department, FAMU-FSU College of Engineering, Florida State University, Tallahassee, Florida 32310, USA}
\address[8]{National High Magnetic Field Laboratory, 1800 East Paul Dirac Drive, Tallahassee, Florida 32310, USA}
\address[9]{Texas A\&M University, Department of Physics and Astronomy, College Station, TX 77843-4242, USA}

\cortext[cor1]{{pratyushkuma@umass.edu}}

\begin{abstract}
We have designed and tested a large-area (0.15~m$^2$) neutron detector based on neutron capture on \ce{^{6}Li}.  The  neutron detector design has been optimized for the purpose of tagging the scattering angle of keV-scale neutrons. These neutron detectors would be employed to calibrate the low-energy ($<$100~eV) nuclear recoil in detectors for dark matter and coherent elastic neutrino nucleus scattering (CE$\nu$NS).  We describe the design, construction, and characterization of a prototype. The prototype is designed to have a tagging efficiency of $\sim$25\% at the relevant $\mathcal{O}$(keV) neutron energies, and with a mean capture time of $\sim$17$~\mu$s. The prototype was characterized using a \ce{^{252}Cf} neutron source and agreement with the simulation was observed within a few percent level.  
\end{abstract}


\end{frontmatter}

\section{Introduction} \label{sec:intro}
Many dark matter (DM) models predict that scattering in target materials will be dominated by coherent scattering with the nucleus. Recent advancements in detector technologies and theoretical understanding of DM~\cite{alkhatibLightDarkMatter2020,barakSENSEIDirectDetectionResults2020,ibeMigdalEffectDark2018,kouvarisProbingSubGeVDark2017,kahn2021searches} have pointed out the possibility of low-mass DM in a mass range from keV to GeV, below the long-standing $>$GeV focus of the field. Numerous detector technologies have been proposed to look for sub-GeV DM~\cite{PhysRevD.96.016026,hertelPathDirectDetection2019,KNAPEN2018386}. In this DM mass regime, the energy deposited by DM in a target nucleus is $\mathcal{O}$(eV) or lower.  While detector technologies are rapidly advancing to these low nuclear recoil (NR) thresholds, \textit{calibration} methods capable of producing and tagging $\mathcal{O}$(eV) NR are lagging.  To perform such low-energy NR \textit{in-situ} calibrations via coincidence techniques, one benefits from a source of monoenergetic neutrons with keV energies and a backing detector capable of measuring the scattering angle of such neutrons in the target.  Here we focus on this backing detector portion of a future $\mathcal{O}$(eV) NR calibration setup.

Neutrons of the keV energy scale are particularly challenging to detect because their energy is typically too small to efficiently cause detectable scintillation from scattering and also typically too large to be efficiently or quickly captured in capture-based neutron detectors. A significant amount of moderation is required before neutron capture can occur.  The drawbacks of this moderate-then-capture detection method are the long capture time (roughly 10$\,\mu$s) and the large radial diffusion distance during moderation (roughly 10$\,$cm)~\cite{YEN2000476}.

Many DM sensor technologies under consideration in this sub-GeV regime rely on phonon propagation and are therefore relatively slow in response.  The risk of pile-up therefore limits the maximum calibration rate and implies NR calibrations of such technologies are naturally statistics-limited.  The neutron backing detector must then have a maximized detection efficiency, meaning both a large solid angle coverage and a large detection efficiency per solid angle.  While detectors with the ability to tag keV neutrons do exist, most of them are either prohibitively expensive per unit area (implying only small angular coverages are practical), or are inefficient (meaning a large fraction of neutrons escape before capture). Table~[\ref{tab:my_label}] shows a few representative neutron detector options in this energy regime. The long-term goal of $\mathcal{O}$(eV) NR calibration would benefit from a high-efficiency capture-based backing detector technology that is also inexpensive (enabling high angular coverage). 

\begin{table}[]
    \centering
\begin{tabular}{ |p{4cm}|p{2cm}|p{3cm}|p{4cm}|   }
\hline
\multicolumn{4}{|c|}{Neutron Detector for $\sim\mathcal{O}$(keV) neutrons } \\
\hline
Detector & Max Efficiency & Geometry & Notes\\
\hline
\ce{^{6}Li}I(Eu) crystal~\cite{barbeauDesignCharacterizationNeutron2007a} & 29\%  at 24keV&2.5 cm radius x 7.5 cm long &high efficiency, small surface area, high price\\
\hline
GS20 glass~\cite{kleinNeutronResonanceTransmission2020a} & -  & 2.54 cm radius × 0.5 cm long& small surface area, high price \\
\hline
\ce{^{10}B}-loaded liquid scintillator~\cite{YEN2000476} &71\% at 1keV &  43 cm radius x 4 cm long  &high efficiency, large surface area, extremely costly \\
\hline

\end{tabular}
    \caption{This table shows the currently used neutron detectors to detect keV-scale neutrons }
    \label{tab:my_label}
\end{table}

The principle of detection of our neutron backing detector is as follows: A keV-scale neutron that has first scattered in a target material is then moderated in the hydrogen-rich backing detector materials (HDPE and acrylic). The moderated neutron is then captured in a thin layer of \ce{^{6}Li}-containing ZnS(Ag) scintillator. The ZnS(Ag) scintillation light propagates into the clear acrylic and is captured in an array of embedded wavelength-shifting (WLS) fibers. Finally, the concentrated light within the fibers is detected using commercial silicon photomultipliers (SiPMs). This article is organized as follows: section~[\ref{sec:Design Studies}] discusses the designing studies, section~[\ref{sec:prototype}] is focused on the assembly and characterization of the detector, and finally we conclude in section~[\ref{sec:conclusions}].

\begin{figure}[h] 
\centering
\includegraphics[width=0.9\textwidth]{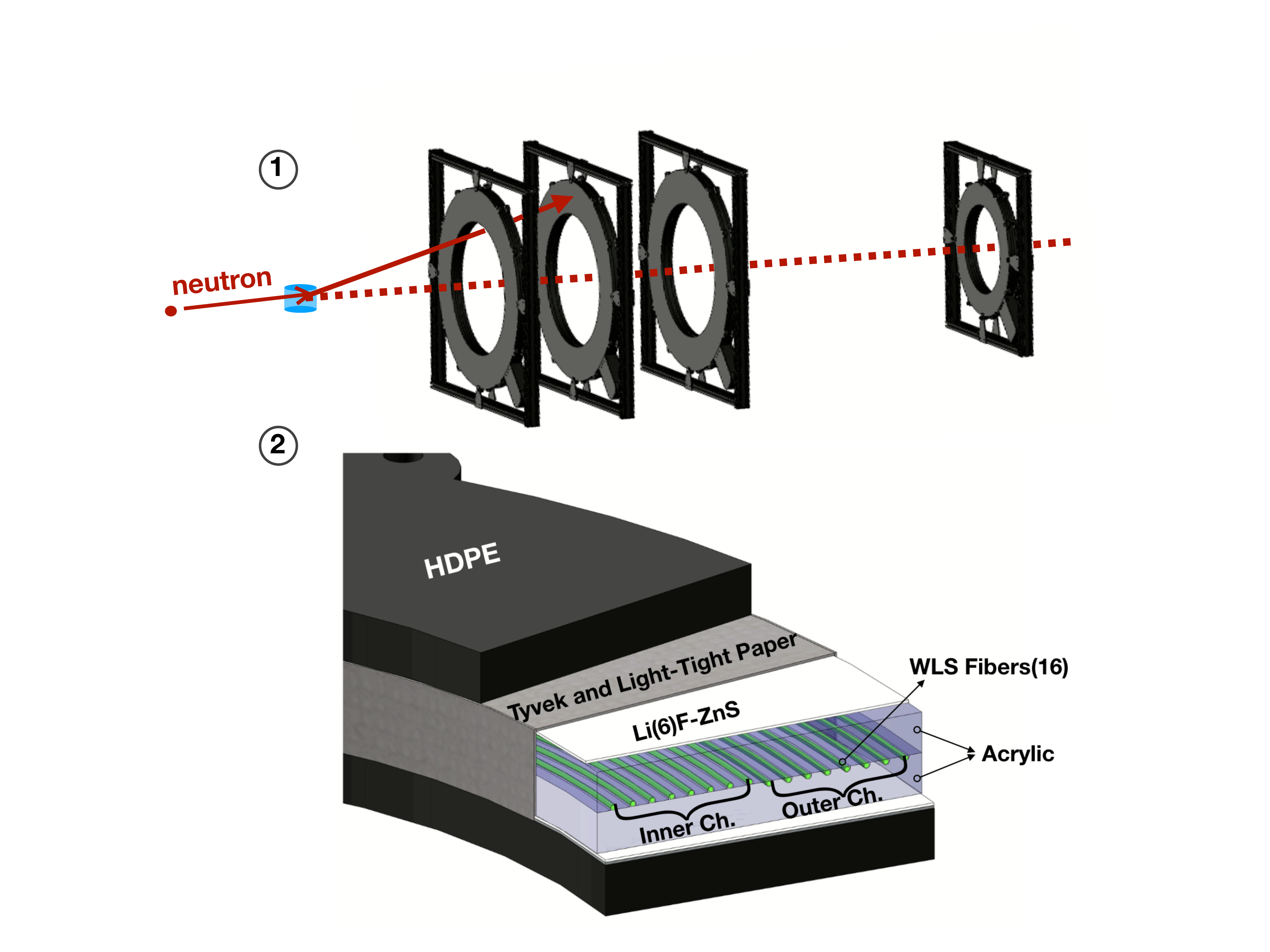}
    \caption{(Top)  Shows  the schematic representation of the calibration setup with the arrays of backing detectors. (Bottom) Shows the cross-sectional view of the backing detector. The central acrylic layer is divided into two parts, and the fibers were embedded on the grooves made on these parts.}\label{fig:backing_detector_crossection}
\end{figure}

\section{DesignStudies}
\label{sec:Design Studies}
The following section describes the goals of the neutron backing detector, which have dictated the various design choices made.
\begin{itemize}

    \item \textbf{Maximum neutron capture efficiency}: Employing a pulsed neutron source ~\cite{CHERNIKOVA201474} to perform \textit{in-situ} calibrations of DM detectors implies that the calibration setup would be limited by statistics due to the limited number of pulses available from a pulsed neutron source. This motivates a large-area neutron detector with a high capture efficiency per neutron.

    \item \textbf{Maximum angular resolution:} Angular resolution is the second most severe limitation to the planned neutron scattering experiments, limiting the ultimate energy precision of the calibration. Angular resolution is challenging in capture based detectors because the moderation process takes place over $\sim$10~cm length scales.  A further degradation in angular measurement occurs when a neutron initially scatters in a backing detector at one angle but escapes and captures in a backing detector at another \emph{incorrect} angle. One can think of this as neutron cross-talk. The dimensions and spacing of the rings must be optimized to maximize both the angular resolution and minimize this neutron cross talk effect while retaining capture efficiency.
    
    \item \textbf{Minimum neutron capture timescale}: Because the planned calibrations are dependent on a coincidence between the pulsed neutron source and the detection of a scattered neutron, the capture timescale would ideally be similar to or less than the source’s pulse duration (in our case, 1~$\mu$s).  The design goal is to minimize the neutron capture time so that the effectiveness of the coincidence timing technique is maximized.  Achieving the 1~$\mu$s capture time goal is challenging at keV energies.  \ce{^{10}B}-doped liquid scintillator can achieve the goal, but with a downside of significant gamma rates and poor gamma rejection via PSD.

    \item \textbf{Minimum contamination from background gammas:} The pulsed neutron source will produce a simultaneous pulse of gammas, many of which are produced by neutron capture on various shielding or incidental materials. The interaction rates of these gammas in the active scintillating material must be minimized if the neutron capture signal is to be observed.  Additionally, it is highly beneficial if the scintillation process distinguishes between gamma scatters and neutron captures. Also, this difference in gamma or neutron signals is benefited by maximum light detection efficiency.

\end{itemize} 

Based on the above constraints, we designed a ring shaped backing detector where the scintillators were sandwiched between moderating hydrogenous materials (Fig.~\ref{fig:backing_detector_crossection}).  A very thin scintillator of thickness 0.5$\,$mm was chosen in order to minimize gamma interactions.  Two layers of scintillating material were used to increase the capture efficiency. Backgrounds originating in the WLS fibers themselves were reduced by using two arrays of fibers to read out the capture signals, requiring coincidence between the two channels read out by separate fiber arrays and SiPMs.  The WLS fibers in the inner part of the detector ring were attached to one of two SiPMs. This comprised the \textit{inner channel}. Similarly, the outer ring WLS fibers and their corresponding SiPM comprise the \textit{outer channel}. The chosen scintillator was EJ-426(ZnS:Ag/\ce{^{6}Li}F)~\cite{wilhelmDevelopmentOperation6LiF2017}, a  95\%-enriched \ce{^{6}Li} solid-state scintillator material in the form of  \ce{^{6}Li}F  dispersed in a ZnS:(Ag) scintillator matrix. The neutron is detected via the n-capture signal on \ce{^{6}Li},  n-capture reaction \ce{^{6}Li}(n, $\alpha$)T produces alpha($\alpha$) and tritium(T)  where the kinetic energy of the $\alpha$ particle is 2.05 MeV, the kinetic energy of T is 2.73 MeV. ZnS(Ag) scintillator matrix has a light yield of about 50,000 Photons/MeV~\cite{yehuda-zadaOptimization6LiFZnS2018}, thus a large light output of 1.6$\,\times\,$10$^5$ photons~\cite{osovizkyDesignUltrathinCold2018} is produces per n-capture signal in the ZnS(Ag) scintillator. ZnS(Ag) also exhibits Pulse Shape Discrimination (PSD) for additional rejection power.  The main downside of ZnS(Ag)/\ce{^{6}Li}F is its poor light transmittance to its own scintillation light due to its structure as a matrix of multiple powdered crystals. In order to ensure neutron capture events from all depths within the ZnS(Ag) are detectable, a high light detection efficiency is still motivated, as will be discussed later.

After selecting the active scintillating material, the next step was to optimize the moderator dimensions so they could efficiently moderate keV scale neutrons to the meV scale required for capture. GEANT4~\cite{AGOSTINELLI2003250} was used for neutron simulations and a custom physics list was created. The Shielding Physics list~\cite{G4_physics} was chosen to simulate neutrons from the thermal range up to an energy of 20$\,$MeV. For thermal energy, G4NeutronThermalScattering was used with a thermal treatment of hydrogen.
4$\,$eV was set as both the maximum energy for thermal scattering and the minimum energy for elastic scattering. There has been good agreement between MCNP and GEANT4 using the thermal scattering data that has been studied by different groups~\cite{vanderendeUseGEANT4Vs2016}.

\subsection{Moderator Layers and Thicknesses}
\label{sec:sim_thicknesses}

HDPE and acrylic were both used as moderators in the detector due to their high hydrogen content and low cost. Simulations were performed to optimize the thickness of these two materials. Optimum thicknesses were found between the extreme of too-thick (the neutron will capture on hydrogen, before reaching the \ce{^{6}Li}-doped scintillator) and too-thin (the neutron will fail to moderate before reaching the scintillator). In addition to finding the combination of HDPE and acrylic thicknesses which maximize capture efficiency, the capture timescale was studied, and some compromise was made between the conflicting goals of high capture efficiency and high timing resolution. Timing resolution is benefited by thinner moderating materials. 
Thus we must find the optimal combination of capture efficiency and capture time.

\begin{figure}[h] 
\centering
\includegraphics[width=\textwidth]{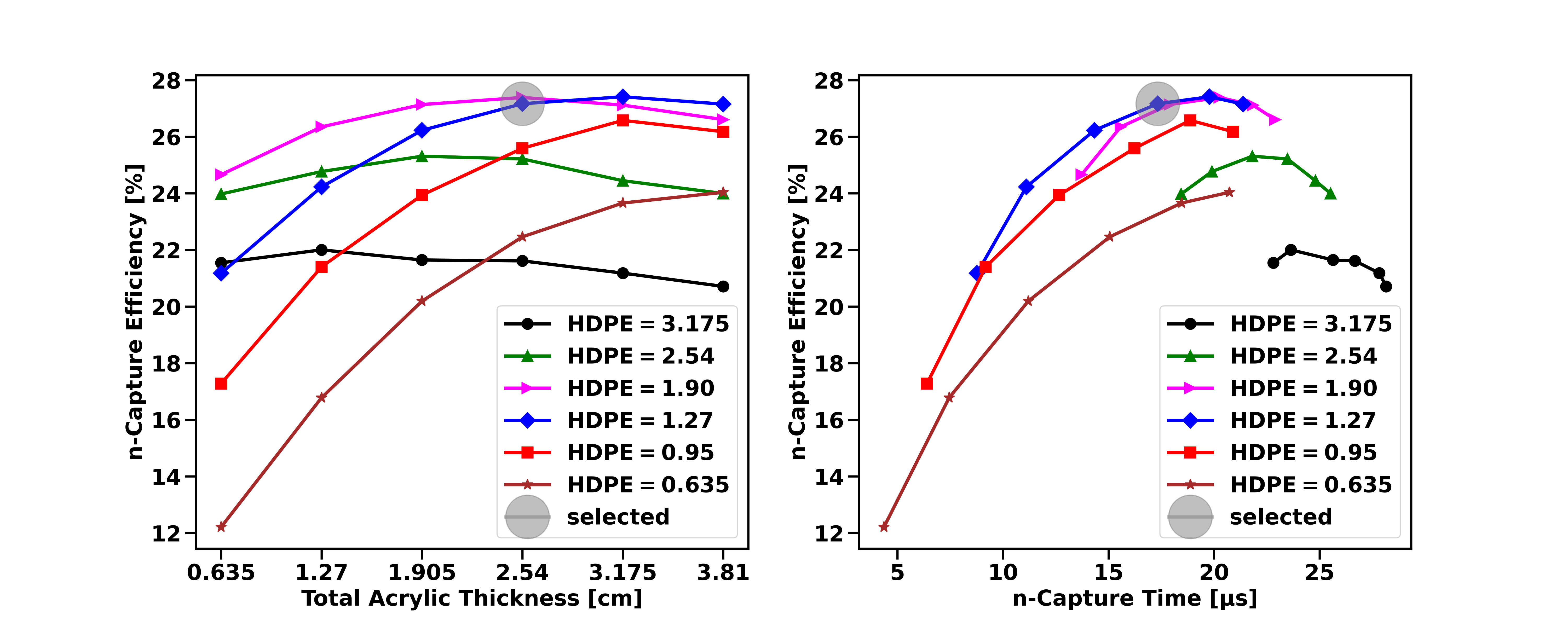}
 \caption{All the dimensions in the legends are in cm. (Left) Shows various capture efficiencies for different thickness of central acrylic layer and HDPE thicknesses. (Right) Shows the corresponding mean capture times for various configurations. The grey dots signify the selected configurations. The selected configuration of the moderator consists of two HDPE layers each 0.5$\,$inch (12.7$\,$mm) thick and a central acrylic layer with a thickness of 1.0$\,$inch (25.4$\,$mm), which together produce a neutron capture efficiency of 27\% at 2$\,$keV and a capture time of roughly 17$\,\mu$s. }\label{fig:optimised graph}
\end{figure}

Results of these thickness-varying simulations are shown in Fig.~\ref{fig:optimised graph}, assuming a 2~keV neutron energy incident perpendicular to the moderating layers. The simulated geometry consisted of a sandwich geometry of HDPE-Acrylic-HDPE layers. The thickness of each layer was varied in the simulations. Based on these simulations, the thickness of each HDPE layer was set to 0.5-inch (12.7~mm) and the thickness of the central acrylic layer was set to 1.0-inch (25.4~mm), with a capture efficiency of 27\% at 2keV.  This is slightly lower efficiency than the maximum achievable (at slightly increased acrylic thickness) but was chosen to reduce the mean neutron capture time to 17$\,\mu$s.

\subsection{Ring Dimensions and Spacing}
\label{sec:sim_light}
After optimizing the acrylic and HDPE thickness for capture efficiency and timing, other aspects of the ring geometry were optimized for angular resolution and minimizing neutron crosstalk between rings. As previously discussed, a neutron will diffuse radially during thermalization before capture, both limiting the position resolution of the detector and risking neutron capture in an adjacent detector. 
To estimate the rough scale of the radial diffusion length for keV neutrons, a histogram of radial distance of the n-capture position was made using the data from the previous simulation. And it was found that most neutrons gets captured after diffusing radially a distance of 8$\,$cm. Thus a ring detector with radial thickness of 15$\,$cm.

\begin{figure}[H] 
\centering
\includegraphics[width=0.9\textwidth]{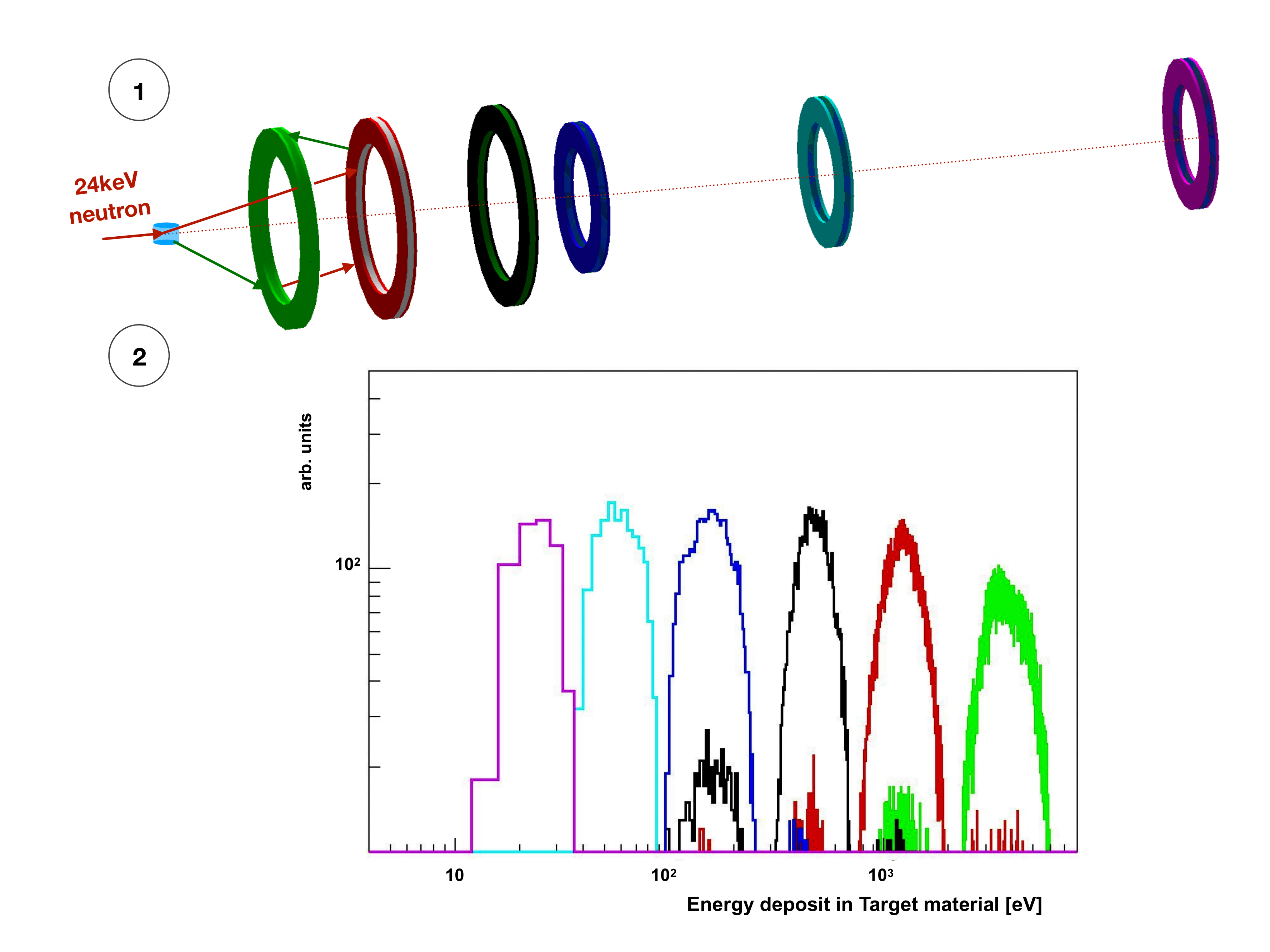}
\caption{(Top) Shows the geometry of the setup that was simulated, array of six backing detectors with the target material and a pencil beam of 24$\,$keV neutrons.  (Bottom) Shows the degree of crosstalk resulting from the configuration in the Top. The scattered 24$\,$keV neutron from the target is supposed to be detected by the second ring, but it scatters off the second ring and gets detected in the first ring instead.}\label{fig:NeutronCrossTalk_effect}
\end{figure}

However, increasing the radial thickness of the detector also increases the probability that neutrons are captured in the wrong ring in the assembly (neutron crosstalk). To study this effect, the full geometry consisting of a small target material, a series of neutron detector rings and a central monochromatic collimated neutron beam shown in Fig.~\ref{fig:NeutronCrossTalk_effect} was simulated. It was observed that neutron crosstalk adds only a minor contamination as long as the spacing between rings is significantly larger than the radial thickness of each ring. In other words, the solid angle coverage of another ring as viewed from the primary ring can be kept low, while the solid angle coverage of the collection of rings is kept collectively high. The simulations showed that crosstalk events safely consisted of less than 10\% of all neutron capture events,  requiring no further fine-tuning or optimization of the geometry.

\begin{figure}[H] 
\centering
\includegraphics[width=0.9\textwidth]{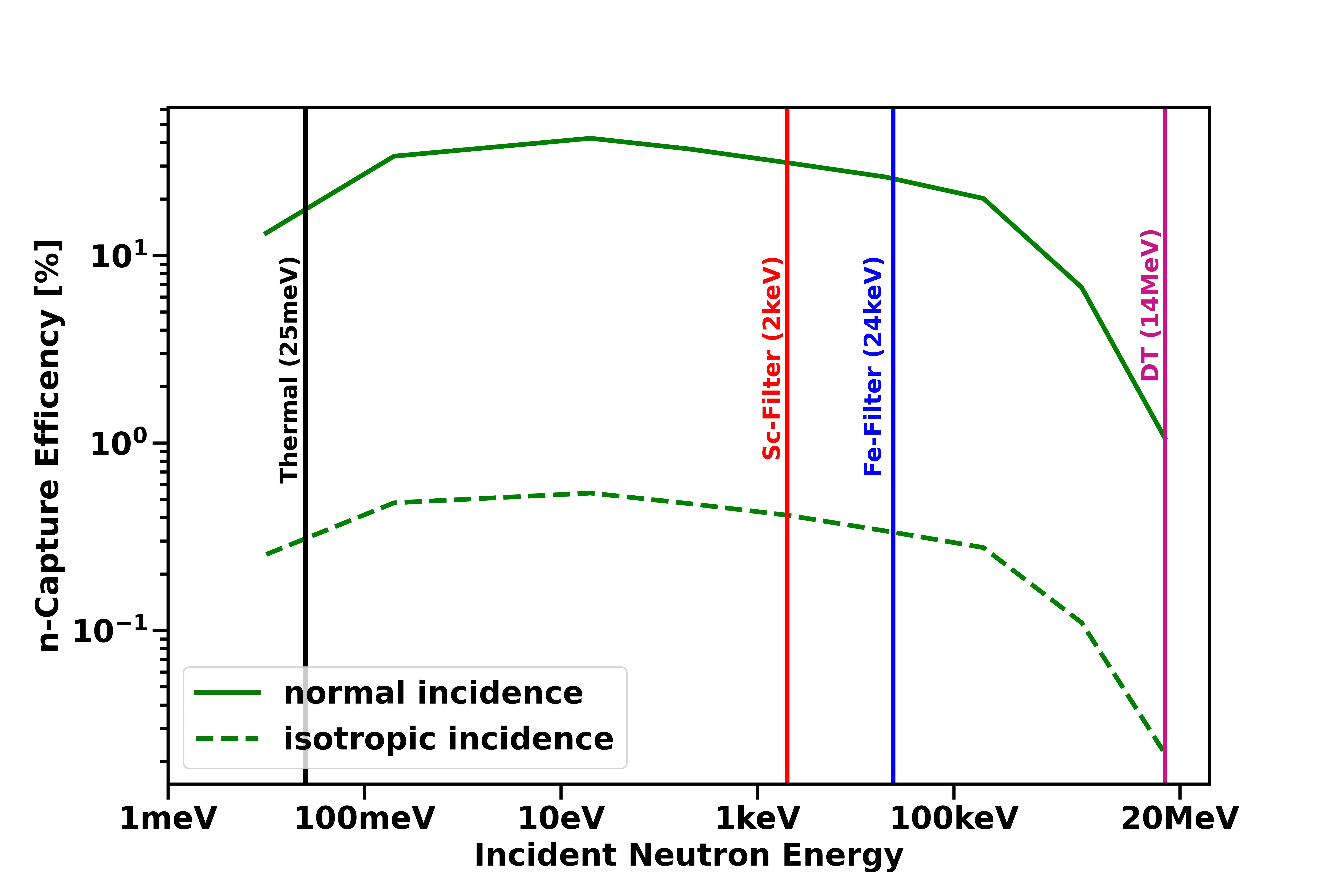}
\caption{Two sets of simulations were performed here: one with neutrons normally incident to the backing detector the other with neutrons isotropically incident to it. The region of interest shows 2$\,$keV and 24$\,$keV neutrons from a Sc and Fe filtered neutron source. Thermal and DT neutrons, the main backgrounds in the NR calibration experiment, are also highlighted. 
}\label{fig:efficiency_to_all_energy_scale}
\end{figure}

From this study, we also estimated the range of scattering angles that are accessible, 10$^{\circ}$ - 60$^{\circ}$  of scattering angle. Due to the relatively large radial thickness of the detector, careful study of the geometry was necessary in order to achieve the desired angular resolution. Finally, to get a holistic picture of the n-capture efficiency of the backing detector across all energy scales, simulations were performed and the results are shown in Fig.~\ref{fig:efficiency_to_all_energy_scale}.  The thermal and DT neutrons would be the main background sources in a NR-calibration experiment based on a filtered neutron source generated with a DT neutron generator. Consequently, the tapering of n-capture efficiency to lower values at the thermal and DT neutron energies is promising. 

\subsection{Light Collection}
\label{sec:sim_light_collection}
\begin{figure}[H] 
\centering
\includegraphics[width=0.9\textwidth]{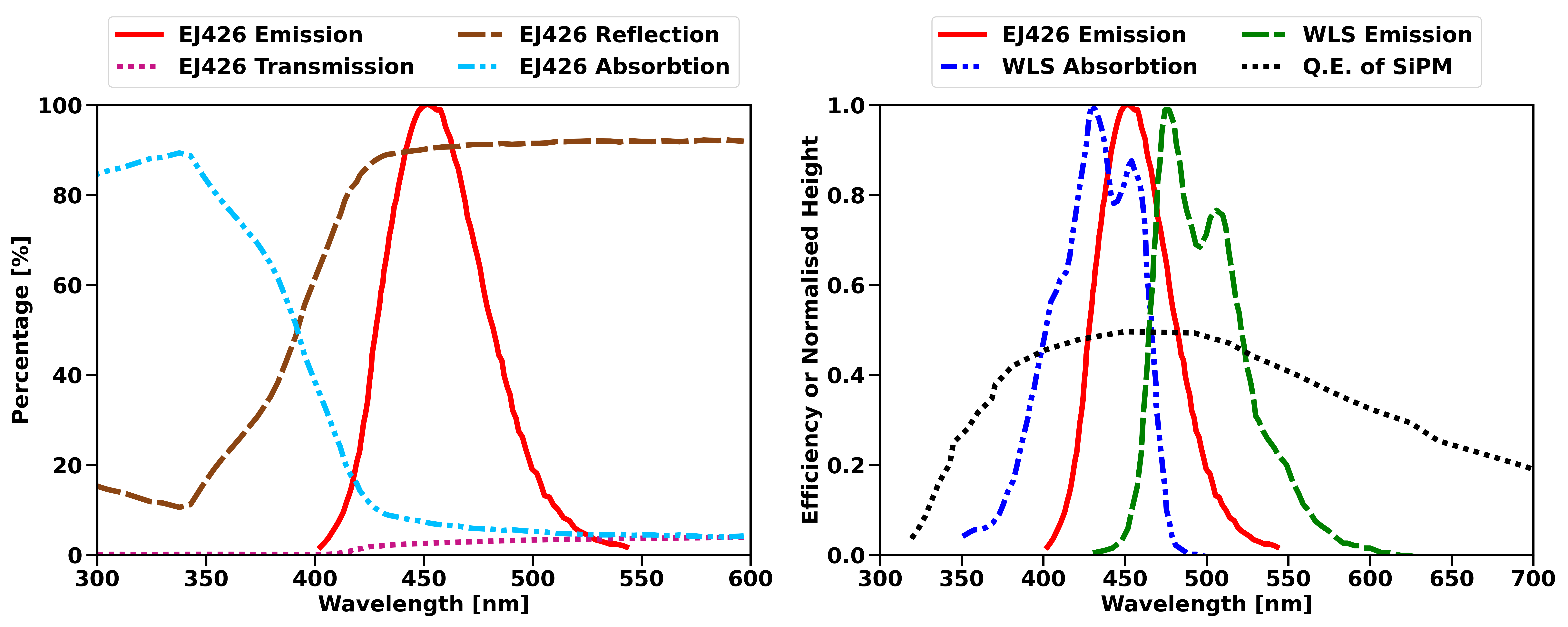}
\caption{(Left) The optical properties of EJ-426 as a function of wavelength, adapted from~\cite{wilhelmDevelopmentOperation6LiF2017}. (Right) The normalized plot of emission and absorption spectra of WLS fibers and the emission spectrum of EJ-426 (\ce{^{6}Li}F-Zns(Ag)) scintillator. Also included is the quantum efficiency of the SiPM. }\label{fig:ANTS2input}
\end{figure}

An efficient optical system is necessary in order to increase the light collection efficiency of the detector (to ensure all neutron captures on \ce{^{6}Li} were above threshold, and to maximize the utility of PSD).  After the scintillation photons escape the  ZnS:(Ag) matrix, they undergo multiple reflections off the acrylic walls aided by a highly reflective Tyvek layer. Then, a large fraction of these photons will be captured within a parallel array of WLS fibers.  We selected a 1.5$\,$mm diameter fiber from Kurray:  Y-11 multicladding non S-type~\cite{abreuOptimisationScintillationLight2018}. These fibers were placed in grooves made on the inner surfaces of the two acrylic sheets. Finally, these WLS fibers were connected to a Hamamatsu S13360 series SiPM~\cite{otteCharacterizationThreeHigh2017} via a connector interface, to be discussed later. The choice of WLS fibers and SiPM was made to ensure a maximum overlap between both the emission spectrum of the \ce{^{6}Li}F/ZnS(Ag) scintillator and the absorption spectrum of the WLS fibers, and the emission spectrum of the WLS fibers and the quantum efficiency spectrum of the SiPM. See  Fig.~\ref{fig:ANTS2input}. 


\begin{figure}[h] 
\centering
\includegraphics[width=\textwidth]{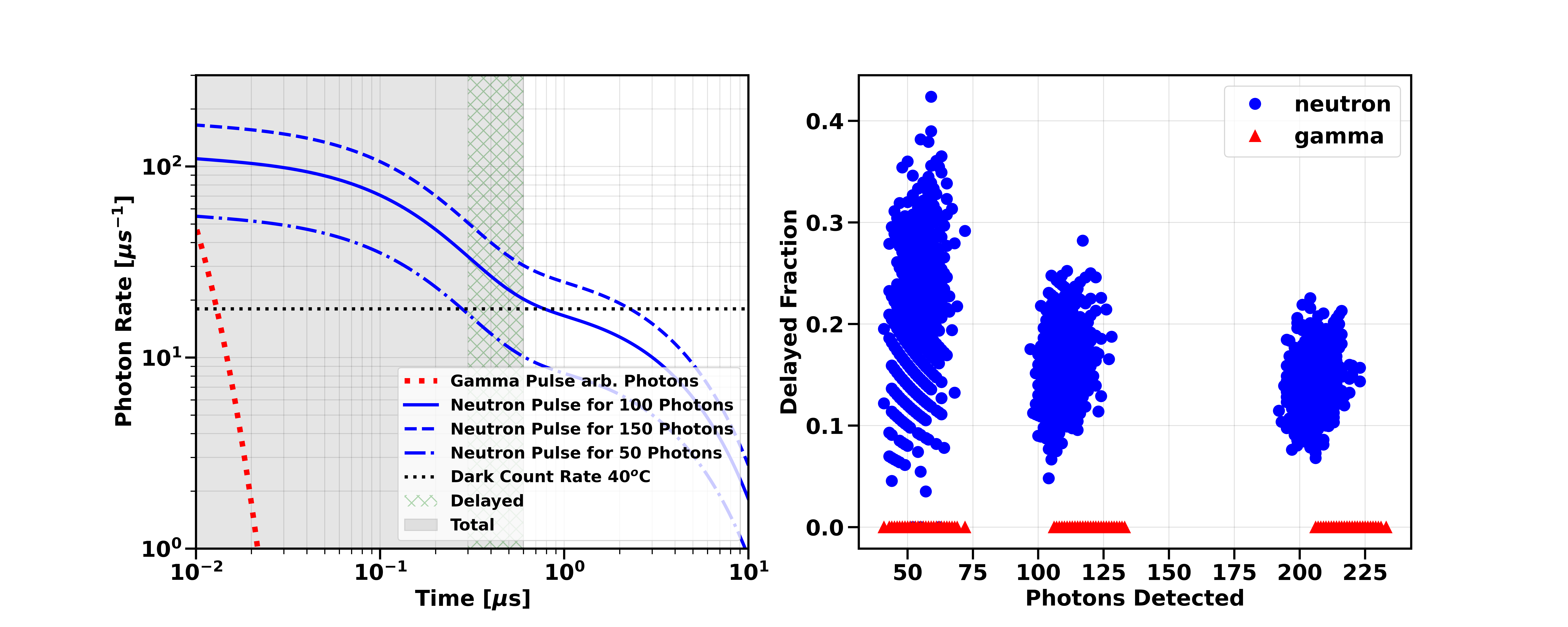}
    \caption{(Left) Typical pulse shapes for a neutron signal (blue) of different sizes and for a gamma signal (magenta), based on~\cite{pinoStudyThermalNeutron2015}. The gamma pulse signal is very short as compared to the neutron signal. (Right) Result of a simple Monte Carlo performed to estimate the lowest number of photons that need to be detected for PSD.}\label{fig:EJ426 Photon Pulse}
\end{figure}

  A Monte Carlo (MC) simulation was performed based on the study by Pino et.al~\cite{pinoStudyThermalNeutron2015}, to estimate the number of photons needed to perform PSD. The n-capture signal from the EJ-426 (\ce{^{6}Li}F-Zns(Ag)) scintillator is of the order of a few $\mu$s, and the gamma signal is of the order of a few ns. Both are highlighted in the left plot of Fig.~\ref{fig:EJ426 Photon Pulse}. This long tail feature in the n-capture signal helps to discriminate between the ER and n-capture events even though the SiPM has a high dark count at room temperature. From the MC simulation it was concluded that capturing roughly 100 photons will be sufficient to discriminate between ER and neutron capture events at room temperature (Fig.~\ref{fig:EJ426 Photon Pulse}, right plot). Next, the number of WLS fibers was determined by performing an optical simulation using ANTS2~\cite{morozovANTS2PackageSimulation2016}. The ability to perform PSD dictated the number of WLS fibers needed. Several inputs to the optical simulation were provided, most important of which were the absorption, transmission and reflection coefficients of the EJ-426 layer (Fig.~\ref{fig:ANTS2input}). The wavelength shifting ability of the WLS fibers was also modeled in the simulations. The refractive indices of various materials as well as their other optical properties were imported from the XCOM: Photon Cross Sections Database~\cite{curtis.supleenist.govXCOMPhotonCross2009}. The simulation suggested that a total of 16 fibers were needed to have $\gtrsim\,$100 photons detected by the SiPMs. 
  
\section{PrototypeFabricationAndInitialTesting}
\label{sec:prototype}

\subsection{Fabrication, Assembly and Readout}
\label{sec:prototype_fabrication}

\begin{figure}[H] 
\centering
\includegraphics[width=0.7\textwidth]{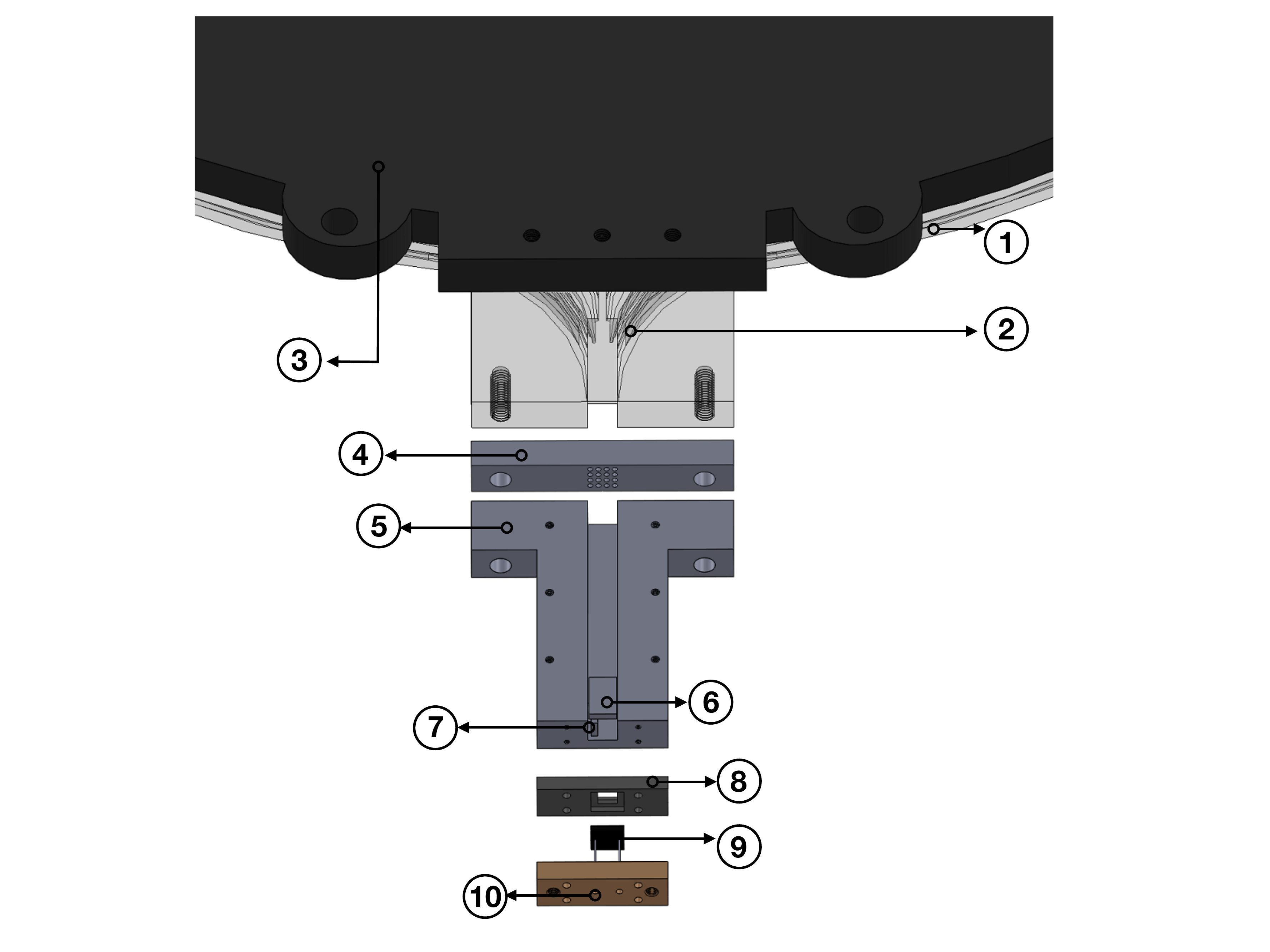}
\caption{\textbf{1)} Grooved acrylic. \textbf{2)} Grooves. \textbf{3)} HDPE for neutron moderation. \textbf{4)} Grid plate to orient the fibers. \textbf{5)}  Tee shaped aluminum slot with upward slanted groove. \textbf{6 \& 7)} Pressing plates. \textbf{8)} Connector interface with chamfered edges. \textbf{9)} SiPM. \textbf{10)} Copper support.
}
\label{fig:clamp_fiber_SiPM}
\end{figure}

 The 16 fibers were grouped into an inner channel and an outer channel. Each channel was then mated to two different SiPMs. The central acrylic layer of 24$\,$mm (optimized thickness was 25.4$\,$mm) was cut into two halves, and each channel was embedded in grooves made in the acrylic. Next, a custom clamping strategy was designed and built to closely pack the fibers together before mating them to the 6$\,$mm$\,\times\,$6$\,$mm SiPM face (see  Fig.~\ref{fig:clamp_fiber_SiPM}). First, the fibers from the acrylic are arranged into a 4$\,\times\,$4 grid with the use of a \textit{grid plate} that helps in orientating the fibers. Next, the fibers are bent inside a tee shaped piece of aluminum, which has an upward slanted groove. Finally, at the other end of the aluminum tee, set screws are used to drive \textit{pressing plates} that hold the fibers in the grid formation. The fibers are then placed through a \textit{connector interface}, which has chamfered edges to provide extra directionality for the outward protruding fibers so that mating with the SiPM is easier. The connector interface also functions as a holder for the SiPM. Once the fibers are fed into the connector interface, they are polished and mated to the SiPM using optical glue. A final copper part was used to secure the SiPM in place and allow the optional use of Peltier coolers. A couple of inner channel fibers suffered slight cracks as a result of this securing process.

 The top and bottom flat surfaces of the acrylic are optically clear, and the grooved portion has a semi mill-finished surface. It was decided that there would be no air gap between the WLS fibers and the grooves in the acrylic. Thus the space between the fibers and grooves was filled with optical glue, and care was taken to minimize the creation of air bubbles as they could act as photon sinks in the system. The different steps of the assembly process are shown below in  Fig.~\ref{fig:Fabrication_process}.
 
\begin{figure}[H] 
\centering
\includegraphics[width=0.8\textwidth]{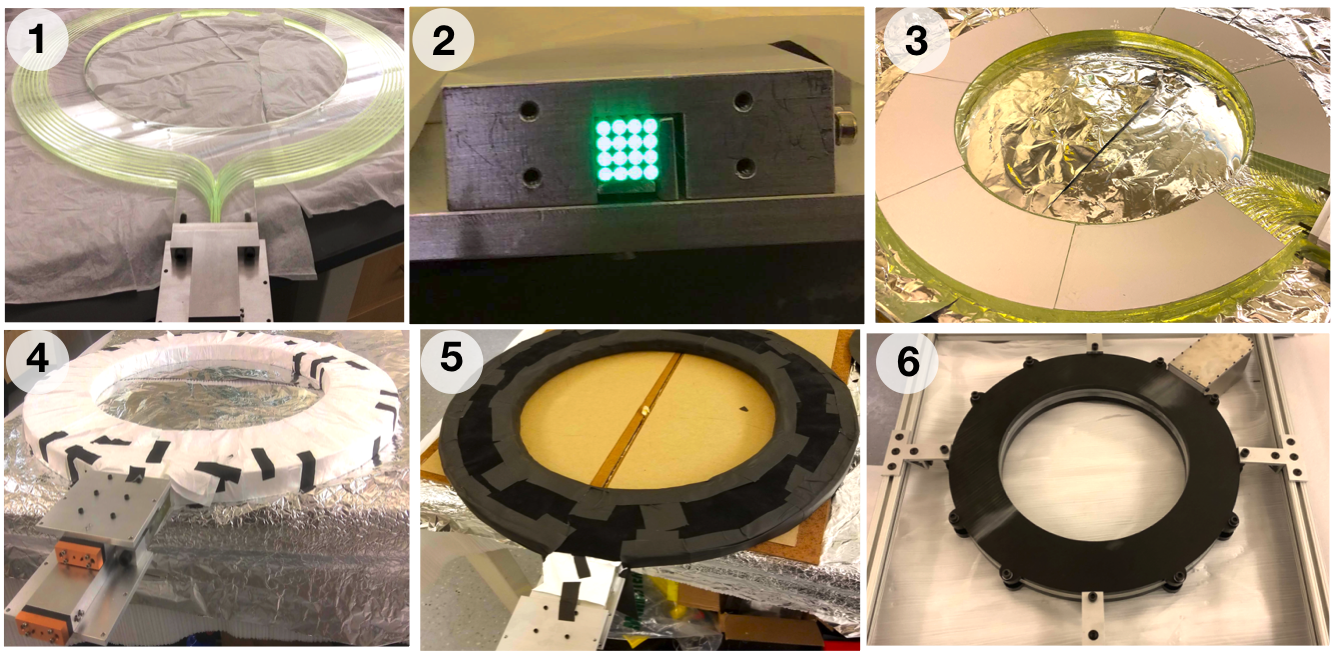}
\caption{
\textbf{1)} Grooved acrylic with all fibers glued into the groove using RTV-615 silicone glue. \textbf{2)} The 4$\times$4 matrix of the fibers coming out of the clamp. \textbf{3)} \ce{^{6}Li}F-Zns(Ag) scintillator attached to the acrylic via optical glue. \textbf{4)} Tyvek is wrapped around the detector for diffusive reflection of light \textbf{5)} A final layer of thick black opaque paper wraps the detector to make it light tight. \textbf{6)} Fully assembled Backing Detector along with its support structure.  
}
\label{fig:Fabrication_process}
\end{figure}

Next we shall discuss the readout strategy of the SiPM. First, the signal from the SiPM was amplified using an off-the-shelf charge sensitive preamplifier from Cremat (CR-113), which has a gain of 1.3$\,$mV/pC. Appropriately biased resistors and capacitors were then connected to the preamplifier circuit. The preamplifier signal was then passed to a shaping amplifier from an ORTEC 671 Spectroscopy Amplifier. An approximate gain of 20 was applied, and the unipolar gaussian output pulse with a pulse width of 500$\,$ns was passed to the Pice676 DAQ. The bipolar output pulse was passed down to build an OR trigger circuit using the signals from both the inner and outer channels, and this OR trigger was used as an external trigger for the Pice676 DAQ. Each triggered event was recorded while operating the Pice676 DAQ at a sampling frequency of 3$\,$MHz, and each sample was 6$\,$µs pre-trigger and 14$\,$µs post-trigger in duration. A custom-made high-pass filter with cutoff frequency 13$\,$kHz was also introduced so as to reduce the unwanted low frequency noise from the recorded unipolar gaussian pulse signal.

\subsection{Analysis}
\label{sec:prototype_SPE}
\begin{figure}[H] 
\centering
\includegraphics[width=\textwidth]{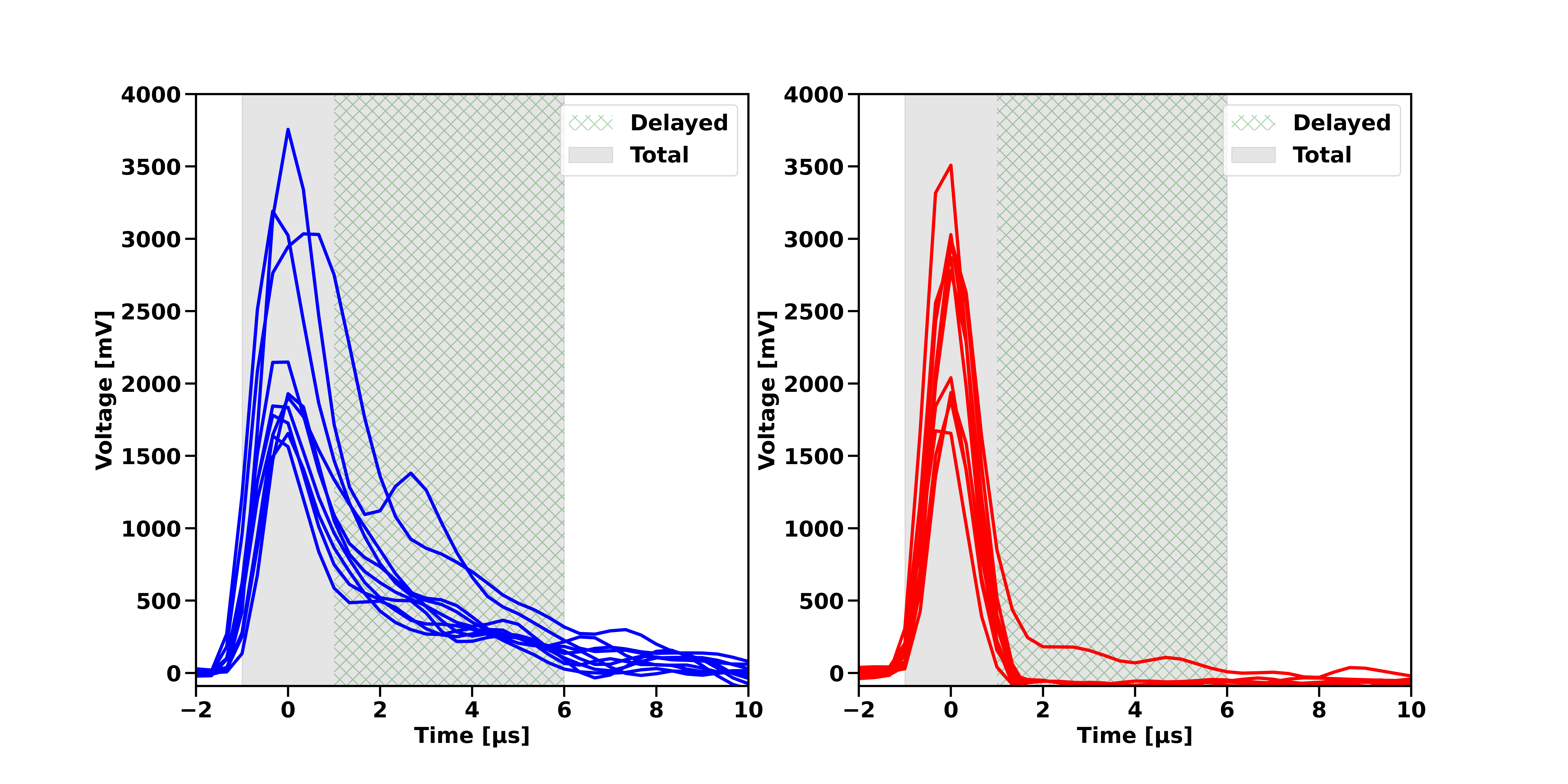}
  \caption{
    This figure shows different types of pulses that are observed with the backing detector. (Left) Neutron-like pulse, scintillation signal caused due to n-capture are relatively slower.
    (Right) Gamma-like pulse, most likely caused by gammas striking the WLS fibers, signals caused due to gamma pulses are relatively faster.
    }\label{fig:different_types_Pulses}
\end{figure}

The experimental data were analyzed to measure the neutron detection efficiency of the neutron backing detector. The primary goal of the analysis process is to clearly distinguish the neutron population from the gamma population. The \ce{^{6}Li}F-ZnS(Ag) has an intrinsic PSD quality based on pulse area as reported in \cite{pinoStudyThermalNeutron2015}. The pulse shape discrimination technique explored in our work is based on the ratio of Delayed Scintillation to that of Total Scintillation. Delayed Scintillation is defined to be the area under the pulse between 1$\,\mu$s and 6$\,\mu$s of the pulse and the total scintillation is defined to be the area under the pulse between -1$\,\mu$s and 6$\,\mu$s, with time zero being the maximum amplitude of the pulse. Since an external hardware trigger was used for this work, a dedicated pulse finder method was not implemented in our analysis. A few pulses are highlighted here in  Fig.~\ref{fig:different_types_Pulses} for illustration.

\begin{equation}
    \text{PSD}=\dfrac{\text{Delayed Scintillation}}{\text{Total Scintillation} }
\end{equation}
 


\begin{figure}[H] 
\centering
\includegraphics[width=\textwidth]{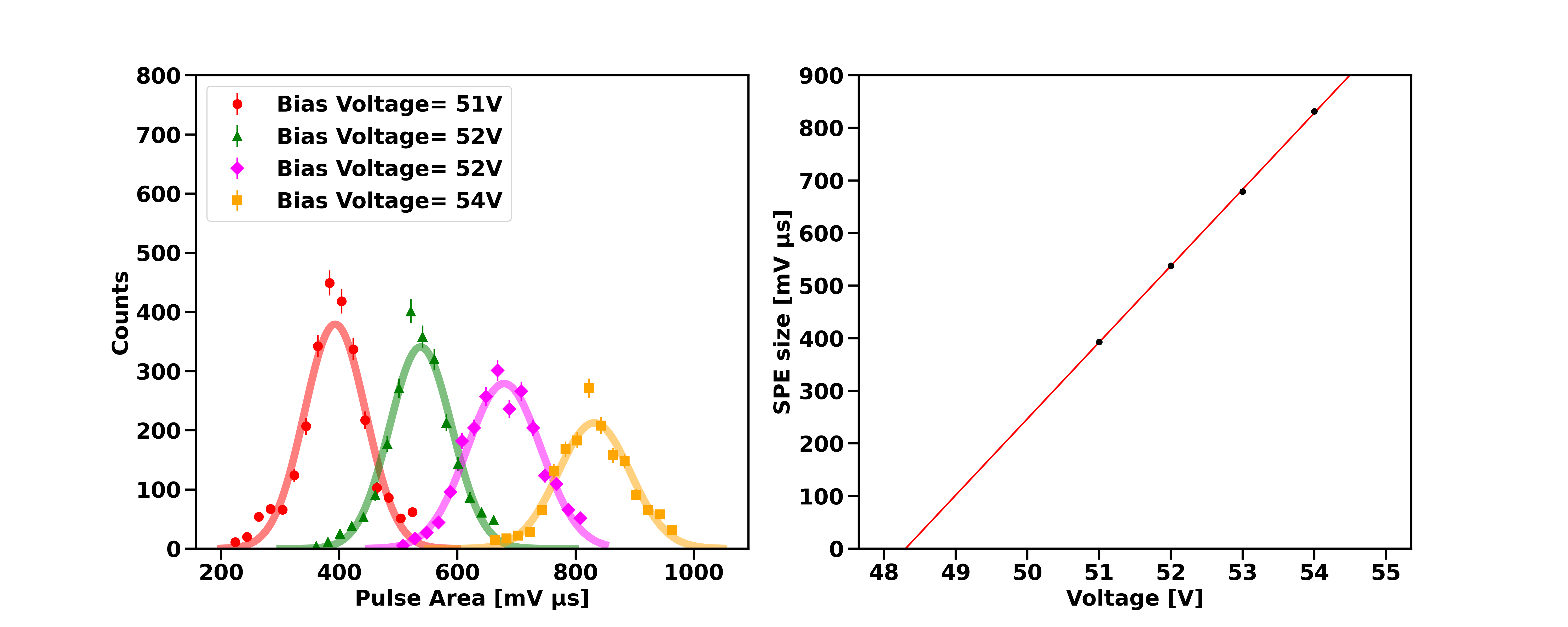}
  \caption{ Measurement was done at dry-ice temperature.
    (Left) SPE sizes at different bias voltages is plotted. SPE value was estimated by performing a gaussian fit at different bias-voltage. 
    (Right) SPE size with varying bias voltage is plotted, from which a linear extrapolation was made to figure out the breakdown voltage of the SiPM. Given this breakdown voltage, an over voltage of 3.1$\,$V was applied and the SPE size was measured. The measured SPE size was assumed to be the SPE size at room temperature. 
}
\label{fig:SPE}
\end{figure}

Signal size was converted to the units of detected photons by dividing the raw pulse area by the measured single photoelectron (SPE) area. Because of a high dark rate at room temperature, the SPE measurement was performed at 194$\,$K using dry ice.  The SPE size is independent of temperature \emph{if} the over-voltage set point (voltage above breakdown voltage) is kept constant~\cite{Nagai:2016vym}.  The breakdown voltage of the SiPM when cold was inferred using the SPE size at different bias voltages (Fig.~\ref{fig:SPE}), and found to be 48$\,\pm\,$0.7$\,$V. At an over-voltage of 3.1$\,$V, the SPE size when cold was found to be 450$\,\pm\,$10$\,$mV*$\mu$s. We then operated the SiPMs at an identical 3.1$\,$V over-voltage at room temperature and assumed the same SPE area. Note: During actual data-taking, the  gain of the amplifier was scaled down by a factor of 40 assuming a linear response of the amplifier circuit. SPE size was also reduced by the same factor.

\subsubsection{Backgrounds}
\label{sec:prototype_backgrounds}
Before describing the final efficiency of the neutron detector, various types of background pulses that need to be rejected are discussed below:  

\begin{itemize}
    \item Fast Pulse (similar to right panel of  Fig.~\ref{fig:different_types_Pulses}): These are purely gamma events as their rate gets higher when gamma sources such as \ce{^{137}Cs} and \ce{^{152}Eu} were placed near the detector. Due to thinness of \ce{^{6}Li}F-ZnS(Ag) scintillator, it  scintillates mostly due to n-capture signal and is blind to most of the gammas. However, a fast pulse can be produced if an ambient gamma ray strikes the WLS fiber causing the fiber to scintillate. Another  cause of these fast pulses is when a high energy gamma from a cosmic ray interacts with the scintillator and deposits considerable energy, though it is very unlikely. The observed rate of such Fast Pulse events is about 10$\,$Hz in ambient room environment.  
    
     \item Slow Pulse in absence of a neutron source (similar to left panel of  Fig.~\ref{fig:different_types_Pulses}): The cause of these backgrounds are likely due to the presence of radioactive materials inside the neutron detector that can decay by the emission of an alpha, which mimics a neutron capture signal to some degree. The rate that has been reported earlier~\cite{kozlovLargeAreaDetector2018a} is quite low (4.88$\times10^{-6}\,cm^{-2}\,sec^{-1}$) as compared to our observation. The rate of Slow Pulse events observed in our backing detector is 0.5$\,$Hz (3.4$\times10^{-4}\,cm^{-2}\,sec^{-1}$).  Another possibility to create a Slow Pulse is due to the capture of ambient neutrons in the environment, but that is also highly unlikely since the \ce{^{6}Li}F-Zns(Ag) layer is covered with HDPE plastic, and any ambient neutrons in the room environment would most likely get captured in the HDPE.    
     
\end{itemize}

\subsubsection{Cuts}
\begin{figure}[H] 
    \centering
    \includegraphics[width=\textwidth]{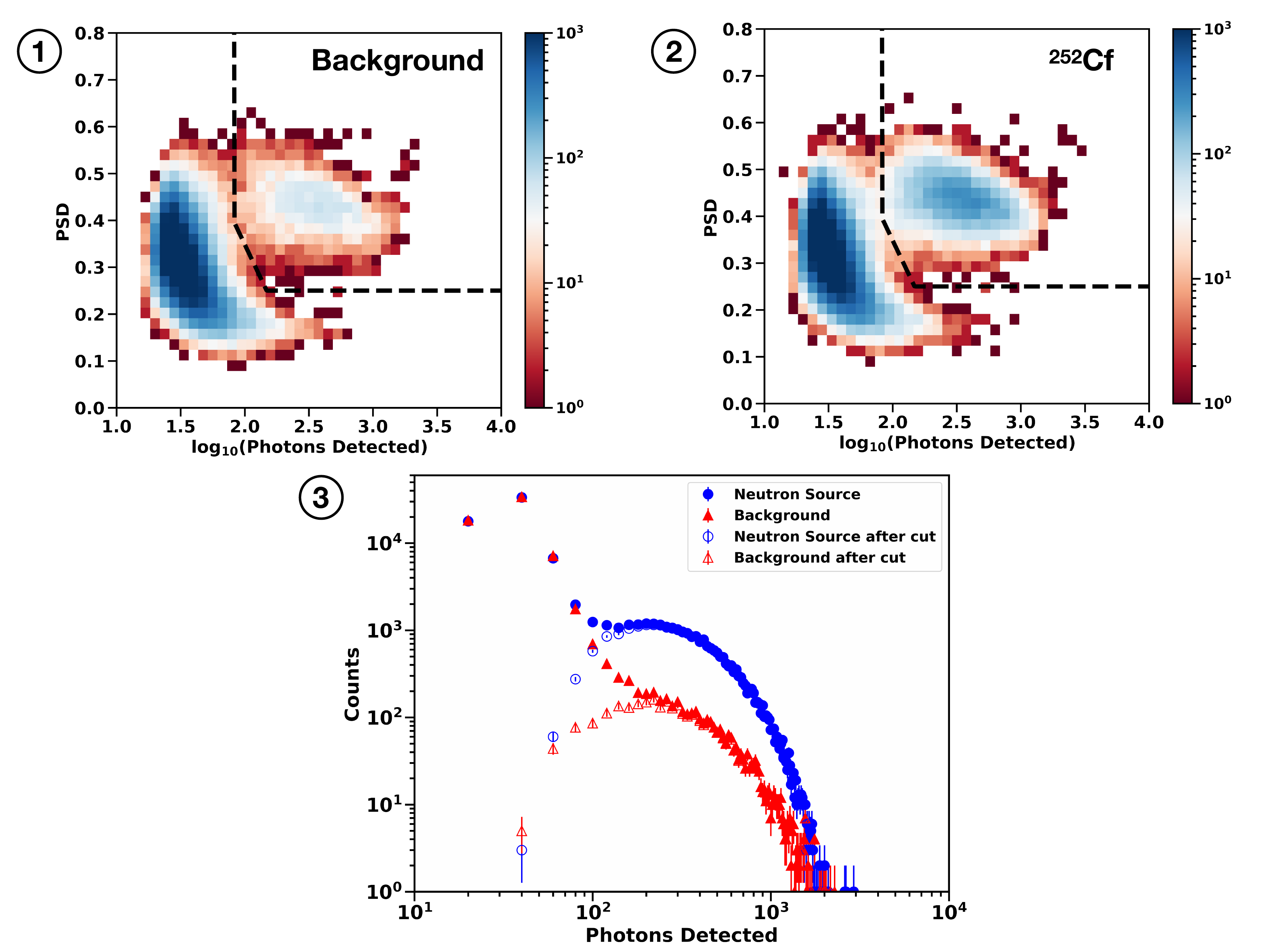}
    \caption{
    The 2-d histogram in the plane of log$_{10}$ of photons detected vs the delayed scintillation obtained after analysis of data from (Panel 1) Background dataset. (Panel 2) from Configuration \textbf{1} dataset (refer Table~\ref{tab:different_geometrical_configuration_table}). The slower and faster populations are clearly distinct in this space. Neutron-like signals are the slower signals and the gamma-like signals are the faster signals. To get the number of n-capture signals, data cuts shown in black dotted line were applied.
    (Panel 3) Shows the photons detected by the detector in presence of the neutron source and in absence of the neutron source. (before and after the data cuts were applied) 
    }
    \label{fig:PSD_cuts}
    \vspace{-0.5em}
    \end{figure}

After defining the Pulse Shape Discrimination variable, finding the true signal size in units of detected photons and having an understanding of the various types of backgrounds, we can finally compare the observed neutron detection efficiency to simulation after applying some data cuts. To highlight the applied data cuts, a plot of PSD parameter vs log$_{10}$(signal size) is shown in  Fig.~\ref{fig:PSD_cuts} for a particular source-detector configuration. One can clearly see the two distinct populations; the slower and the faster. Separation of these two populations in this 2d space was done by performing a combination of linear cuts. The clear separation is a signature that almost all the n-capture signals are being detected without the loss of any light signal. Next we will characterize the level of agreement of the optical model with experiment.

\subsection{Light collection}
\label{sec:prototype_light}
In this section we evaluate the light collection efficiency of the setup.  The number of photons collected from our experimental setup and the simulation was compared. The total light collected matches the simulation if the photons collected from simulation were scaled up by a factor of five. The following are some of the reasons due to which discrepancy can be seen:

\begin{itemize}

    \item\textbf{Imperfect tuning of optical parameters}: 
    Many optical parameters were taken as inputs to the optical simulation, such as reflectivity of Tyvek sheets, reflectivity of EJ-426-HD and the absorption coefficient of the WLS fibers, which may vary in some percent level with each specimen. An extensive tuning of various optical parameters in our simulation setup was not performed. 
    
        \item\textbf{Optical crosstalk}: 
    Given the large size of the light signal, there is a chance that this might cause optical crosstalk among the various pixels of the SiPM and enhance the size of our signal.~\cite{masudaSuppressionOpticalCrosstalk2021}

\end{itemize}

\begin{figure}[H] 
\centering
\includegraphics[width=\textwidth]{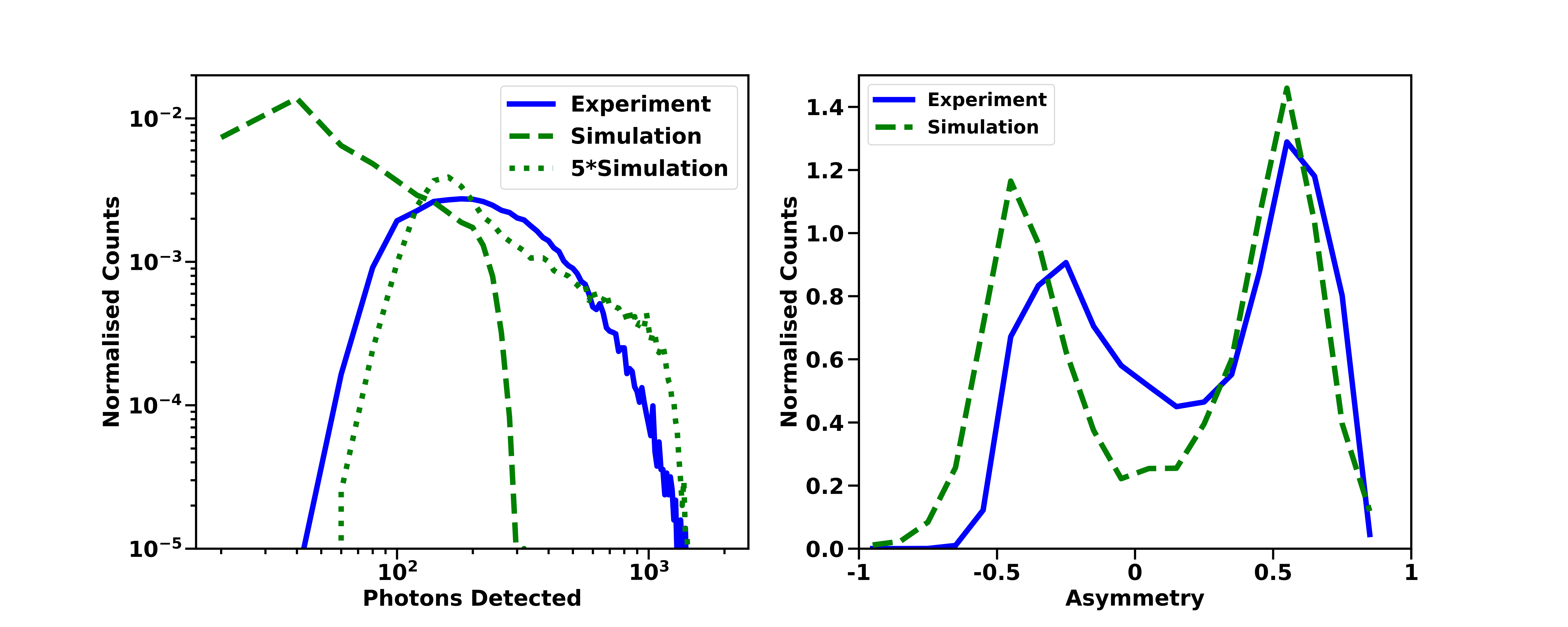}
   \caption{
    (Left) Compares the number of photons detected by the backing detector with the  simulation. A factor of 5 discrepancy between the number of photons detected and simulation was found. Possible reasons for this are discussed in the main text.
    (Right) Asymmetry of the signal matches with optical model, validating our model. Some degree of angular information is also retained in the signal. If needed, a distinction between a capture event happening in the inner channel vs outer channel can be made.
    }\label{fig:Photons_detected}
\end{figure}

  The end goal of the backing detector is to tag keV scale neutrons efficiently; we are not concerned with the amount of energy that is deposited by each n-capture event. Thus correct simulation of light propagation is of greater relative importance. To quantify the agreement of the light propagation model with experiments, we compared the \textit{asymmetry} of the light signals in both the inner and outer channels which is defined as \ref{eq:2}. The right panel of figure~[\ref{fig:Photons_detected}] shows the asymmetry of the light signal, which is in good agreement with simulation. The outer channel has larger amounts of fiber (due to the larger radius of curvature) and thus more photons are captured in it, which is the reason the right peak is larger than the left peak. As discussed earlier, a few of the fibers in the inner channels were slightly damaged. This is reflected in the skewness of the asymmetry towards the outer channel in the data obtained from the experiments. The right panel of figure~[\ref{fig:Photons_detected}] also signifies that some degree of radial information is retained as the capture events happening at the outer edge of the detector will deposit most of the light in the outer channel and vice versa.  
\begin{equation}
    \text{Asymmetry} = \dfrac{\text{Outer channel - Inner channel}}{\text{Outer channel + Inner channel}} \label{eq:2}
\end{equation}

\subsection{Efficiency}
\label{sec:prototype_efficiency}
\begin{figure}[H] 
    \centering
    \includegraphics[width=\textwidth]{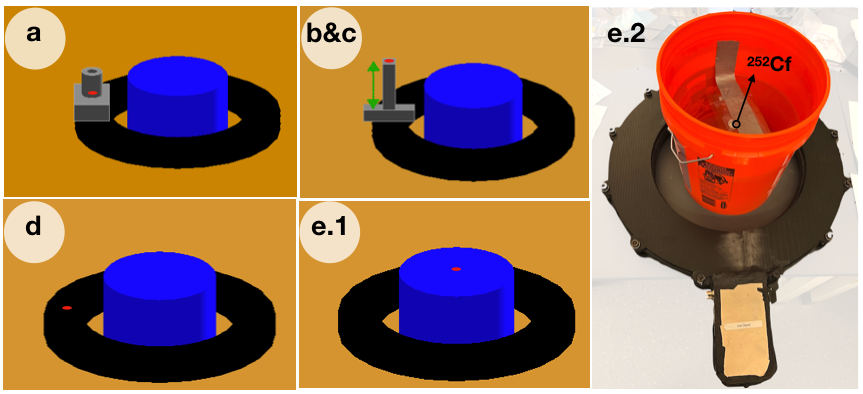}
    \caption{Simulated Geometry of different configurations. The red dot represents the source position. The blue cylinder represents the bucket of water that was placed at the center of the backing detector (black colored cylindrical object). 
    \textbf{a)} Configuration 1 
    \textbf{b, c)} Configuration 2, which comprises two configurations \textbf{b)} Short Al Tee \textbf{c)} Long Al Tee
    \textbf{d)} Configuration 3 
    \textbf{e.1)} Configuration 4
    \textbf{e.2)} Actual setup. 
    }
    \label{fig:different_configuration_image}
    \end{figure}

The absence of low-energy neutron sources at our immediate disposal makes it challenging to measure the neutron detection efficiency of the backing detector for 2$\,$keV and 24$\,$keV neutrons. Validations were instead performed using a non-monoenergetic \ce{^{252}Cf} source in various source-detector configurations as described in Fig.~\ref{fig:different_configuration_image} and Table~\ref{tab:different_geometrical_configuration_table}.  For each configuration, the capture rate in a Geant4 simulation was compared with experimental observation. The simulations used the same physics lists as mentioned earlier.  The \ce{^{252}Cf} source had a neutron yield of 80 neutrons/s. 
  
\begin{figure}[H] 
    \centering
    \includegraphics[width=\textwidth]{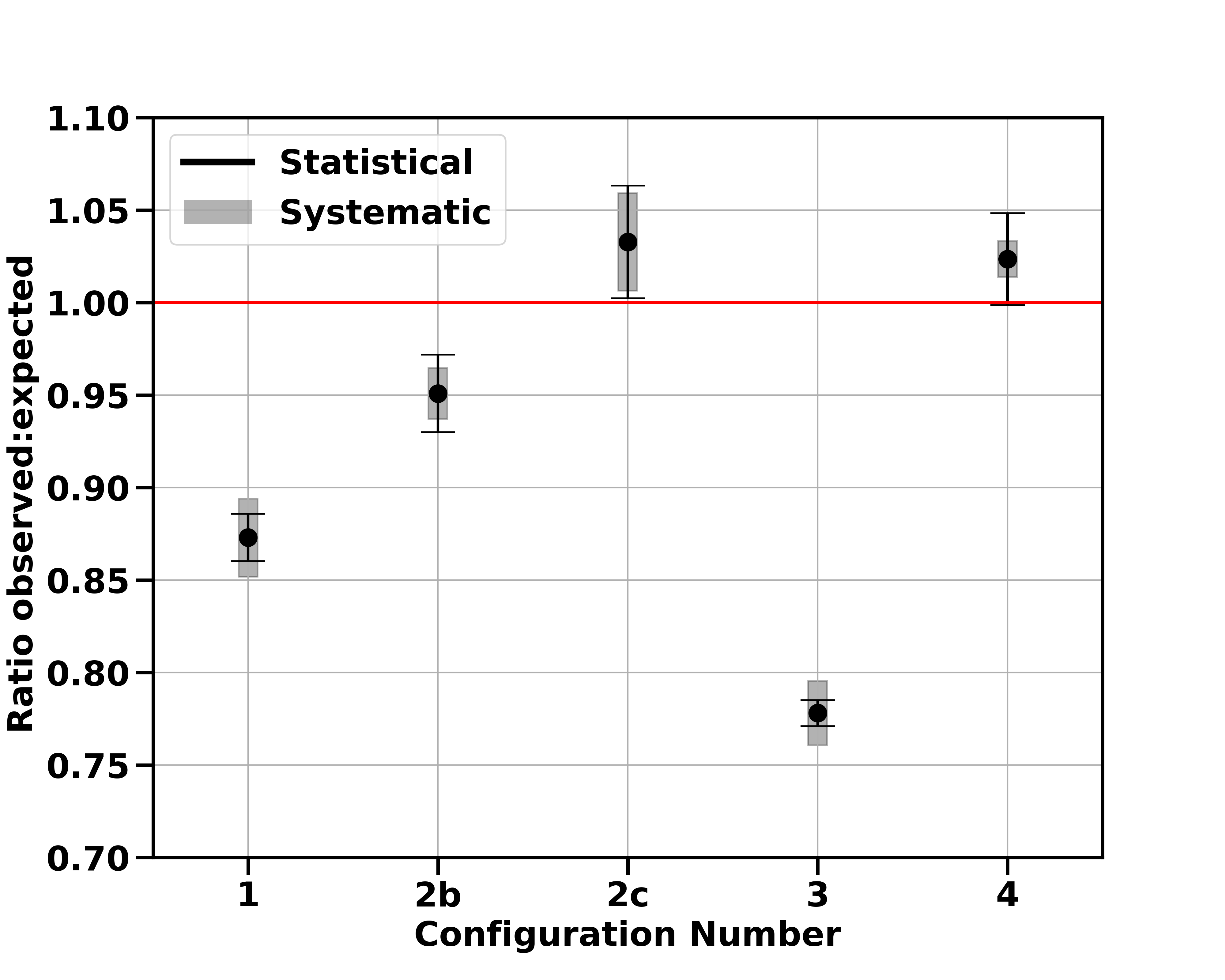}
    \caption{The ratio of observed to expected n-capture signals from the set-up simulated using GEANT4. The solid black lines with caps represent statistical uncertainty. And the gray patch represents systematic uncertainty. Systematic uncertainty were computed using the effect of position error uncertainties, from the source-target detector configuration and density uncertainties,  assuming a uncertainty of 2\% for the density of LiF-ZnS(Ag) Scintillator and the 3\% uncertainty in the activity of the \ce{^{252}Cf} source,a s informed by the manufacturer.}
    \label{fig:Final efficiency}
    \vspace{-0.5em}
    \end{figure}

We define a simple figure of merit: the ratio of observed counts (after background-subtraction) to expected counts given by the simulation.  As seen in Fig.~\ref{fig:Final efficiency}, this relative ratio ranges from 0.78 to 1.03 depending on the configuration (where 1.0 would signify perfect agreement).  This level of agreement is consistent with expectations from the literature.~\cite{gressierInternationalKeyComparison2014}

\begin{table}

    \label{tab:different_geometrical_configuration_table}
\centering
\vskip 3mm
\begin{tabular}{ |p{6cm}|p{6cm}|   } 
\hline
\multicolumn{2}{|c|}{List of various geometrical configurations} \\
\hline
Configuration Number & Description \\
\hline
Configuration \textbf{1} & Source is kept inside a  \\
& lead pig.\\
\hline
Configuration \textbf{2(b)} and \textbf{2(c)}  & Source is kept on an  \\
& Al tee, whose height is varied.\\
\hline
Configuration \textbf{3}  & Source is kept directly \\
& on the detector.\\
\hline
Configuration \textbf{4}  & Source is kept in a bucket  \\
& of water, which is placed\\
& at the center of the detector. \\
\hline
\end{tabular}
    \caption{Shows the different configurations, each labeled with a different number, and a short description of the setup. Refer to Fig.~\ref{fig:different_configuration_image}}
\end{table}

\section{Conclusions}
\label{sec:conclusions}


We have successfully built and characterized a large area backing detector for keV scale neutrons whose neutron detection efficiency matches simulation.  We also demonstrated that some degree of position information is retained by looking at the degree of asymmetry in the light received in both channels. The ambient background rate of our large scale backing detector is 10$\,$Hz, thus the probability of having an accidental coincidence in a window of 10$\,\mu$s, which is the n-capture time window for keV scale neutrons, is negligible. We plan to construct more such backing detectors, which shall be employed to perform low energy nuclear recoil calibrations of detectors for dark matter and coherent elastic neutrino nucleus scattering.      

\section{Acknowledgement}
\label{sec:acknowledgement}

This work was supported by DOE Grant DE-SC020374, and DOE Quantum Information Science Enabled Discovery (QuantISED) for High Energy Physics (KA2401032). This work was completed in part with resources provided by the University of Massachusetts' Green High Performance Computing Cluster (GHPCC). We thank Thomas Langford from Yale University and Jeffrey S. Nico from National Institute of Standards and Technology for useful discussion. W. G. acknowledges the support from the National High Magnetic Field Laboratory, which is supported by National Science Foundation Cooperative Agreement No. DMR-1644779 and the State of Florida. P.K.P. would also like to thanks Chandan Ghosh from University of Massachusetts, Amherst for assistance in improving the triggering logic.

\bibliography{BAckingDetector_Paper_NIMA.bib}


\end{document}